\documentclass[runningheads]{llncs}

\usepackage[inline]{enumitem}
\usepackage[usenames,dvipsnames,svgnames,table]{xcolor} %

\usepackage[utf8]{inputenc}
\usepackage{xargs}
\usepackage{xspace}
\usepackage{url}

\usepackage{rotating}               %

\usepackage{caption}                %
\usepackage{cmll}                   %
\usepackage{colonequals}            %
\usepackage{environ}                %
\usepackage{etex,etoolbox}          %
\usepackage{multirow}               %
\usepackage{nicefrac}

\usepackage{subcaption}             %
\usepackage{thmtools, thm-restate, thm-autoref}  %
\usepackage{xifthen}                %
\usepackage{appendix}               %

\usepackage{stmaryrd}
\usepackage{pdfsync}
\usepackage{graphicx}               %
\usepackage{mathtools} %

\usepackage{hyperref}
\hypersetup{
  colorlinks = true,
  linkcolor = DarkOliveGreen,
  urlcolor  = DarkOliveGreen,
  citecolor = blue,
  anchorcolor = blue
}
\usepackage[capitalise]{cleveref}
\crefformat{enumi}{condition~#2#1#3}
\crefname{fact}{Fact}{Facts}
\Crefname{fact}{Fact}{Facts}
\crefformat{fact}{Fact~#2#1#3}
\DeclareMathSizes{10}{10}{6}{6}

\newcommand{\bnfdef}{\; ::= \;}
\newcommand{\bnfmid}{\;\ \big|\ \;}

\newcommand{\ifempty}[3]{%
  \ifthenelse{\isempty{#1}}{#2}{#3}%
}
\newcommand{\etc}{\textit{etc.}}

\newcommand{\confidence}{\ell}
 
\usepackage{cite}

\usepackage{amsmath}
\usepackage{amssymb}

\usepackage{sessiontypes}

\usepackage{mathtools}

\usepackage{tikz}
\usetikzlibrary{decorations.pathmorphing, calc}
\usetikzlibrary{arrows}
\usetikzlibrary{shadows}
\usetikzlibrary{positioning}

\usepackage{wrapfig}

\usepackage{hyperref}

\usepackage{cleveref}

\usepackage{xspace}

\usepackage{xcolor}
\definecolor{dkgreen}{rgb}{0,0.6,0}
\definecolor{gray}{rgb}{0.5,0.5,0.5}
\definecolor{mauve}{rgb}{0.58,0,0.82}

\usepackage[ruled,vlined,linesnumbered]{algorithm2e}
\Crefname{algocfline}{Algorithm}{Algorithms}

\newcommand{\mon}{\textsf{mon}\xspace}
\newcommand{\synth}{\textsf{synth}\xspace}
\newcommand{\srv}{\textsf{Srv}\xspace}
\newcommand{\client}{\textsf{C}\xspace}
\newcommand{\lchannels}{\texttt{lchannels}\xspace}

\newcommand{\ie}{\textit{i.e.,}\xspace}
\newcommand{\eg}{\textit{e.g.,}\xspace}

\def\finex{{\unskip\nobreak\hfil
\penalty50\hskip1em\null\nobreak\hfil$\diamond$
\parfillskip=0pt\finalhyphendemerits=0\endgraf}}
\newcommand*{\exqed}{\finex}

\definecolor{lime}{HTML}{A6CE39}
\DeclareRobustCommand{\orcidicon}{
	\hspace{-2.5mm}
	\begin{tikzpicture}
	\draw[lime, fill=lime] (0,0) circle [radius=0.16] node[white] {{\fontfamily{qag}\selectfont \tiny ID}};
	\draw[white, fill=white] (-0.06599,0.095) circle [radius=0.007];
	\end{tikzpicture}
	\hspace{-3mm}
}

\foreach \x in {A, ..., Z}{\expandafter\xdef\csname orcid\x\endcsname{\noexpand\href{https://orcid.org/\csname orcidauthor\x\endcsname}{\noexpand\orcidicon}}
}

\begin{document}
\title{Towards Probabilistic Session-Type Monitoring\thanks{%
      The final authenticated version is available online at \url{https://doi.org/10.1007/978-3-030-78142-2_7}.
      The work has been partly supported by: the project MoVeMnt (No:\,217987-051) under the Icelandic Research Fund;
      the BehAPI project funded by the EU H2020 RISE under the Marie Sk{\l}odowska-Curie action (No:\,778233);
      the MIUR projects PRIN 2017FTXR7S IT MATTERS and 2017TWRCNB SEDUCE;
      the EU Horizon 2020 project 830929 \textit{CyberSec4Europe};
      the Danish Industriens Fonds Cyberprogram 2020-0489 \textit{Security-by-Design in Digital Denmark}.
    }
}
\author{Christian Bartolo Burl\`{o}\inst{1}\orcidA{}\and
Adrian Francalanza\inst{2}\orcidB{}\and
Alceste Scalas\inst{3}\orcidC{}\and\\
Catia Trubiani\inst{1}\orcidD{}\and
Emilio Tuosto\inst{1}\orcidE{}
}
\authorrunning{C. Bartolo Burl\`{o} et al.}
\institute{Gran Sasso Science Institute, L'Aquila, Italy\\\and
Department of Computer Science, University of Malta, Msida, Malta\\ \and 
DTU Compute, Technical University of Denmark, Kongens Lyngby, Denmark
}
\maketitle
\begin{abstract}

We present a tool-based approach for the runtime analysis of communicating processes grounded on probabilistic binary session types. 
We synthesise a monitor out of a probabilistic session type where each choice point is augmented with a probability distribution. 
The monitor observes the execution of a process, infers its probabilistic behaviour and issues warnings when the observed behaviour deviates from the one specified by the probabilistic session type.     
\end{abstract}

\section{Introduction}\label{sec:introduction}

Communication is  central to present day computation.
The expected communication protocol between two parties can be formalised as a \emph{(binary) session type}, typically describing qualitative aspects such as the order and choice of service invocations at their corresponding payloads.
In recent work, \emph{quantitative} aspects of the communication protocol are also layered over a session type~\cite{imptt20,DBLP:journals/corr/abs-2011-09037}. 

\begin{example}
  \label{ex:intro}
  Consider a server hosting a guessing game by selecting an integer $n$
  between 1 and 100.
  A client can repeatedly
  \begin{enumerate*}[label=\textit{(\roman*)}]
    \item try to guess,
    \item ask for a hint, or
    \item quit the game.
  \end{enumerate*} 
  The expected interaction sequence of the guessing game server can be
  specified with the session type $\stS[game]$ below:
  \begin{align*}
    \stS[game] = \stRec{X}.\stBraOp
    \left\{\begin{array}{l}
          \stRcv{Guess}{\stVarDec{num}{\typeInt}}{0.75}.
          \stSelOp\left\{\begin{array}{l}
                    \stSnd{Correct}{}{0.01}.\stRecVar{X},
                    \\
                    \stSnd{Incorrect}{}{0.99}.\stRecVar{X}
                    \end{array}\right\},
          \vspace{.2em}\\
          \stRcv{Help}{}{0.2}.\stSnd{Hint}{\stVarDec{info}{\typeStr}}{1}.\stRecVar{X},
          \vspace{.2em}\\
          \stRcv{Quit}{}{0.05}.\stEnd \end{array}\right\}
  \end{align*}
  The server waits for the client's choice (at the external branching point $\stBraOp$) to either \lab{Guess} a number, ask for \lab{Help}, or  
  \lab{Quit}.
  If the client asks for help then the server replies with a
  \lab{Hint} message including a string and the session loops.
  After the outcome of a guess, (described by the internal choice
  $\stSelOp$) is communicated to the client, the protocol recurs.
  The type for the client is \emph{dual} and denoted by
  $\overline{\stS[game]}$: each \lq$\stSelOp$\rq\ is swapped with
  \lq$\stBraOp$\rq\ and each \lq$\stSyn{!}$\rq\ is swapped with
  \lq$\stSyn{?}$\rq.

Besides enforcing that parties follow a certain communication
pattern, our (augmented) session types also specify some quantitative aspects of the
protocol.
For instance, $\stS[game]$ above specifies that the server should give
the client a realistic chance of guessing correctly (\ie $1\%$), while
$\overline{\stS[game]}$ specifies that the client should request for help $20\%$ of the time.
According to this augmented specification with quantitative requirements, a burst of client \lab{Help} requests without any attempts to
guess the correct answer (that far exceeds $20\%$ of total requests)
would constitute a violation of the protocol. 
More specifically, a substantial deviation from the expected behaviour could be seen as an indicator of abnormal behaviours such as an attempted denial-of-service attack.
\exqed
\end{example}

Session types (and their probabilistic variants) are usually checked statically, by type checking the code of the interacting parties (\eg the clients and the server in this case).
However, in an open network, it is common for one or more interacting parties \emph{not} to be available for analysis in a classical pre-deployment fashion. 
There are even cases where, although we have full access to the participants, it is hard to statically verify their behaviour (\eg when a client is a human being, or generated via machine-learning techniques).
This forces the verifier to carry out certain correctness checks 
in a post-deployment phase of software production.
Recent work has shown that detections and enforcements of qualitative process properties expressed in terms of automata-like formalisms can be carried out effectively, at runtime, using monitors~\cite{DBLP:journals/tcs/BocchiCDHY17,AcetoAFIL19,FMT20,DBLP:conf/forte/BurloFS20,DBLP:conf/ecoop/BurloFS21}. 
There are however limits to a monitoring approach for verification \cite{Monitorability:17,Monitorability:21}. 
At runtime, a monitor 
\begin{enumerate*}[label=\textit{(\roman*)}]
  \item%
    cannot observe more than one execution,
  \item%
    can only observe finite prefixes of a (possibly infinite) complete execution, and
  \item%
    cannot detect execution branches that could never have been taken (\ie the analysis is evidence-based). 
\end{enumerate*} 
These constraints make it unclear whether quantitative behavioural aspects such as the observation of branching probabilities, defined over complete  (\ie potentially infinite) sequences of interactions, can be adequately monitored at runtime. 
Note that, in order to determine (with absolute certainty) whether the client-server interactions will observe the probabilities prescribed in $\stS[game]$ above at runtime, one needs access to the code for \emph{both} the server \emph{and} the client. 
Having access to just the source code of the server---as it is reasonable to expect when the specification is dictated by the party providing the service---does not help us in determining whether the probabilities at the external branching point $\stBraOp$ will be observed, since these depend on the choices made by the client (which is often determined at runtime).

In this paper, we develop a tool-supported methodology for the runtime monitoring of
quantitative behaviour for two interacting components.
In particular, given a Probabilistic Session Type (PST) such as
$\stS[game]$ above, we synthesise a monitor that, at runtime, 
\begin{enumerate*}[label=\textit{(\roman*)}]
\item observes the messages exchanged within the protocol to ensure that they follow the protocol prescribed by the session type, 
  \item estimates the probabilistic behaviour of the interacting parties from the interactions observed at runtime, and 
  \item determines whether %
    to issue a warning for behavioural deviations from the branching probabilities prescribed in the PST, up to a pre-specified level of confidence.
\end{enumerate*}
For generality, we target the scenario with the weakest level of assumptions, namely where the monitor and its instrumentation are oblivious to the actual implementation of \emph{both} interacting parties participating in the session. 
Nevertheless, our solution still applies to cases where we have access to the source of the interacting parties.

The rest of the paper is structured as follows.
\Cref{sec:methodology} explains our methodology in detail. 
This lays the necessary foundations for the construction of our monitoring tool, described in \Cref{sec:tool}.
Comparisons to related work and discussions about possible future work are given in \Cref{sec:discussion}.
To the best of our knowledge, the work we present here is the first attempt at verifying PSTs via runtime monitoring. 

\section{Methodology}
\label{sec:methodology}

Our proposed methodology operates post-deployment, where \emph{both} participants in the session are analysed \emph{at runtime}. 
As mentioned earlier, to maximise its applicability and generality, our methodology does not require any part of the participants' behaviour to be analysed pre-deployment, effectively treating them as \emph{black boxes}. 
The definite verdict of whether an execution exhibited by a system abides by some probabilistic specification can only be given once it terminates (if at all). 
Nonetheless, our methodology is able to issue probabilistic judgements from incomplete executions, based on the interactions observed up to that point.

Our runtime analysis employs \emph{online} passive monitors~\cite{Schneider:2000,Monitors:21}.
These are computational entities that run live and observe the \emph{incremental}  
behaviour of two communicating parties (as the execution proceeds) without affecting their interactions.
Although our monitors have no prior information about the actual behaviour of the two
parties,   
they are nevertheless able to: 
\begin{enumerate}[label=\emph{(\alph*)}]
\item\label{item:insight:estimate}%
  approximate on-the-fly the probabilistic behaviour of the interacting parties, %
  iteratively revising the approximation when a new interaction is observed, and
\item\label{item:insight:confidence}%
  make (revocable) judgements that are based on the \emph{probability distribution} described by our PSTs, for a preset \emph{confidence level}. 
\end{enumerate}
Our methodology is supported by a tool, discussed in \Cref{sec:tool}, that automatically synthesises a monitor from a PST $\stS$. 
At run-time, the monitor estimates the probabilities for each choice point of $\stS$ (by observing the messages being sent and received), and determines whether such estimates respect the desired probabilities specified in $\stS$. 
This way, the monitor can apportion blame to the interacting party 
that has control at the choice point where a potential violation is detected.
In the sequel, we present the technical details of our methodology;
in \cref{sec:discussion} we discuss possible alternatives.

\subsection{Probabilistic session types (PSTs)}
\label{sec:prob-session-types}
In order to formalise probabilistic protocols, we adopt session types
augmented with probability distributions over the choice points
($\stBraOp$ and $\stSelOp$): they allow us to specify the probability
of a particular choice being taken by one of the components
interacting in a session.
The syntax of our PSTs (from which $\stS[game]$ in
\Cref{sec:introduction} is derived) is:
\begin{align*}
  \stS \;\bnfdef\; &\ \stBrai{i}{I} && \textit{(external choice)} \\
     \bnfmid & \stSeli{i}{I} && \textit{(internal choice)} \\
  \bnfmid & \stRec{X}.\stS && \textit{(recursion)} \\ 
  \bnfmid & \stRecVar{X} && \textit{(recursion variable)} \\
  \bnfmid & \stEnd && \textit{(termination)}
\end{align*}
In choice points ($\stBraOp$ and $\stSelOp$) the indexing set $I$ is finite and non-empty,
the \emph{choice labels} $\texttt{l}_i$ are pairwise distinct, and the
\emph{sorts} $\asort_i$ range over basic data types ($\typeInt$,
$\typeStr$, $\typeBool$,\etc).
Every choice point $\stS[j]$ is given a \emph{multinomial distribution} interpretation. 
We assume that $\sum_{i \in I} p_i = 1$ where every $p_i$ is positive,
and represents the probability of selecting the branch labelled by $\texttt{l}_i$,  over \emph{every} choice point of interest $\stS[j]$.
The probabilities prescribed at a choice point impose a behavioural obligation on the interacting party who has control over the selection at that choice point. 
For instance, at the external choice in \Cref{ex:intro}, it is the client that is required to adhere to the probabilities prescribed.
As usual, we assume that recursion is guarded, \ie a recursion variable $X$ can only appear under an external or internal prefix.

\subsection{Monitoring sessions}
For the sake of the presentation, we assume that choice points of a PST $\stS$ are indexed by a finite set of indices $j \in J$, which allows us to uniquely identify each choice point as $\stS[j]$. 
Accordingly, we let $I_j$ be a set
indexing the labels $\texttt{l}_{i,j}$ of the choice point $\stS[j]$, and denote the probability assigned at $\stS[j]$ to the branch
labelled by $\texttt{l}_{i,j}$ as $p_{i,j}$.
Our runtime analysis maintains the following counters:
\begin{itemize}
\item $c_j$: number of times the choice point $\stS[j]$ is observed at run-time;
\item $c_{i,j}$: number of times the label $\texttt{l}_{i,j}$
  ($i \in I_j$) of choice $\stS[j]$ is taken.
\end{itemize}
For each $j \in J$, these counters yield the \emph{estimated probability}:
\begin{align}\label{eqn:prob-fun}
  \widehat{p_{i,j}} \;=\; \dfrac{c_{i,j}}{c_j}
  \qquad
  j \in J,\; i \in I_j
\end{align}
Namely, $\widehat{p_{i,j}}$ is the frequency with which the $i$-th branch $\texttt{l}_{i,j}$ of choice point $\stS[j]$ has been taken \emph{so far}.
The monitor continuously updates the estimated probabilities as it observes the interactions taking place while the execution unfolds. 

\newcommand{\SE}{{S}\!{E}}

The monitor cannot base its decision to issue a warning only on these estimated probabilities. 
These could potentially be very inaccurate if either of the components briefly exhibits sporadic behaviour at any point in time of execution. 
To assess whether the monitored sequence of interactions has \emph{substantially} deviated from the probabilistic behaviour specified in a session type, the runtime analysis needs to consider how accurate these estimated probabilities are in conveying the observed behaviour of the components. 
We relate this problem to \emph{statistical inference}, where the sequence of interactions observed up to the current point of execution is a \emph{sample} of the larger population, being the entire (possibly infinite) execution.

There are various established paradigms for statistical inference. 
Our proposed methodology takes a \emph{frequentist} approach. 
In particular, we calculate \emph{confidence intervals} (CIs) \cite{Newcombe12} around each desired probability $p_{i,j}$ in a session type to give an approximation of the expected probabilistic behaviour based on the sample size and a \emph{confidence level} $0 \leq \confidence < 1$. 
For any $\stS$-abiding execution that iterates through choice point $\stS[j]$ for $c_j$ times, the interval would contain the acceptable range of values for the estimated probabilities with confidence $\confidence$. 
To calculate the CI for a choice point $\stS[j]$, we first calculate the \emph{standard error} $\SE$\ on the specified probabilities $p_{i,j}$, which depends on the number of times that choice point $c_{j}$ has iterated (\ie the sample size). 
This is then used to calculate the \emph{maximum acceptable error} $E$ \eqref{eqn:err-fun} based on the given confidence level $\confidence$, where the multiplier $Z(\confidence)$ is the number of standard deviations of a normal distribution representing the particular branch, covering $(\confidence \times 100)\%$ of its values~\cite{Newcombe12}.
\begin{align}
  \label{eqn:err-fun}%
  E_{i,j} \;=\; Z(\confidence)\cdot \SE_{i,j}
  \qquad
  \text{where } \SE_{i,j} \;=\; \sqrt{\dfrac{p_{i,j}(1-p_{i,j})}{c_j}}
  \quad
  \text{for } j \in J,\; i \in I_j
\end{align}
Having calculated the error $E_{i,j}$, the runtime analysis calculates the confidence interval around $p_{i,j}$ as: 
\begin{align}
  \label{eqn:CI}
  & \left[p_{i,j} - E_{i,j},\,\; p_{i,j} + E_{i,j} \right]  
\end{align}
If and when an estimated probability $\widehat{p_{i,j}}$ \eqref{eqn:prob-fun} falls
\emph{outside} this interval, the proposed runtime analysis for our
methodology issues a \emph{warning} implying that:

\begin{quotation}\em
  \noindent
  The estimated probability $\widehat{p_{i,j}}$ has deviated enough from the specified
  probability $p_{i,j}$ to conclude, with confidence $\confidence$,
  that the interacting party responsible for the choice point
  $\stS[j]$ violates the prescribed probability.
\end{quotation}

The higher the confidence level $\confidence$ specified, the longer it takes for the maximum error $E_{i,j}$ in \eqref{eqn:err-fun} to converge~\cite{Newcombe12}.
Consequently:
\begin{itemize}
  \item when a higher confidence $\confidence$ is required, the monitor will have \emph{wider} confidence intervals, hence it needs to collect more evidence (\ie a larger sample size) in order to issue a warning;
  \item when a lower confidence $\confidence$ is required, the monitor will have \emph{narrower} confidence intervals, hence it might deem an observed session to deviate substantially from the probabilities specified in $\stS$ at an earlier point in execution. This means that the monitor may potentially issue spurious warnings.
\end{itemize}
Importantly, after a warning is issued, the subsequent behaviour of the monitored components might cause
the (updated) estimated probabilities to fall back within the confidence intervals. 
As a result, the monitor may \emph{retract} the warning. 
The warnings issued by the monitor become irrevocable verdicts only when the session terminates (if at all). 

\begin{example}\label{ex:monitoring-estimates}
  Recall $\stS[game]$ from \Cref{ex:intro}. Assume that a monitor for $\stS[game]$ is instantiated with confidence level $\confidence = 99.999\%$,
  and that in the running session, the client's choice ($\stBraOp$) has iterated nine times, with the client choosing \lab{Help} five times.
  Thus, the runtime analysis counters are:

  \smallskip\centerline{\(
    c_{\stBraOp} = 9 \qquad\qquad c_{\lab{Help},\stBraOp} = 5
  \)}\smallskip

  \noindent%
  The monitor calculates the estimated probability $\widehat{p_{\lab{Help},\stBraOp}} = 0.56$ from these counter values using \eqref{eqn:prob-fun}. 
  It then calculates the error $E_{\lab{Help},\stBraOp} = 0.59$ from \eqref{eqn:err-fun} (with $Z(\confidence) = 4.4172$) for the $\lab{Help}$ branch in $\stS$. 
  Using the specified probability $p_{\lab{Help},\stBraOp} = 0.2$ and $E_{\lab{Help},\stBraOp}$, it calculates the confidence interval from \eqref{eqn:CI} $0.2 \pm 0.59$, that is $[-0.39, 0.79]$. 
  Since the estimated probability $\widehat{p_{\lab{Help},\stBraOp}}= 0.56$ %
  falls within this confidence interval, 
  the monitor does \emph{not} issue a warning. 

  Now, assume that the session continues, the external choice point \stBraOp \hspace{0.2 mm} is iterated, and the client chooses \lab{Help} eight consecutive times more. 
  This means that the counters of our runtime analysis become:

  \smallskip\centerline{\(
    c_{\stBraOp} = 17 \qquad\qquad c_{\lab{Help},\stBraOp} = 13
  \)}\smallskip

  From \eqref{eqn:prob-fun},  the estimated probability is updated to $\widehat{p_{\lab{Help},\stBraOp}} = 0.76$.
  From \eqref{eqn:err-fun}, the monitor also updates the confidence interval.
  Since $c_{\stBraOp}$ is now larger, it yields $E_{\lab{Help},\stBraOp} = 0.43$ which results in the narrower confidence interval $[-0.23, 0.63]$. 
  At this point, our runtime analysis detects that the estimated $\widehat{p_{\lab{Help},\stBraOp}}$ falls outside this confidence interval and the corresponding warning is issued.
  Note that a monitor for $\stS[game]$ with lower confidence level $\confidence = 95\%$ (\ie $Z(\confidence) = 1.9599$) would issue a warning earlier, \eg when $c_{\stBraOp} = 9$ and $c_{\lab{Help},\stBraOp} = 5$. 
  In fact, the lower confidence level would yield $E_{\lab{Help},\stBraOp} = 0.26$ giving the tighter confidence interval $[-0.06,0.46]$ which does not include $\widehat{p_{\lab{Help},\stBraOp}} = 0.56$. \exqed
\end{example}

\section{The Tool}
\label{sec:tool}

We extend the monitoring framework in~\cite{DBLP:conf/forte/BurloFS20} to implement our probabilistic session type monitors.
The implementation is available at:%

\begin{center}
  \url{https://github.com/chrisbartoloburlo/stmonitor}
\end{center}
%
%
%
\begin{wrapfigure}[4]{c}{3.8cm}
  \centering
  \vspace{-0.7cm}
\definecolor{red}{gray}{0.9}
\centering
\begin{tikzpicture}
  \node (pa) [fill=red, draw=black, rounded corners, align=center] {\footnotesize\srv};
  \node (Lconnmanager) [draw=gray, text=gray, align=center, right=0.25cm of pa, inner sep=1pt] {\scriptsize$\textsf{CM}$};
  \node (mon) [densely dashed, draw=black, align=center, right=0.1cm of Lconnmanager, inner sep=4pt] {\footnotesize\mon};
  \node (Rconnmanager) [draw=gray, text=gray, align=center, right=0.1cm of mon, inner sep=1pt] {\scriptsize$\textsf{CM}$};
  \node (pb) [fill=red, draw=black, rounded corners, align=center, right=0.25cm of Rconnmanager] {\footnotesize\client};
  \node (synth) [densely dotted, draw=black, fill=white, align=center, above=0.3 of mon] {\footnotesize$\textsf{synth}$};
  \node (sessiontype) [draw=white, fill=white, align=center, text height=0.2cm, text=black, outer sep=0pt, left=0.4 of synth] {\footnotesize$\stS$};

  \draw[thin,double distance=1.5pt] (pa) -- (Lconnmanager);
  \draw[thin,double distance=1.5pt] (Rconnmanager) -- (pb);
  \draw[thick, draw=gray][-] (mon) edge (Rconnmanager);
  \draw[thick, draw=gray][-] (mon) edge (Lconnmanager);
  \draw[-angle 90] (synth) edge (mon);
  \draw[-stealth] (sessiontype) -- (synth);
\end{tikzpicture}
   \vspace{-0.7cm}
\end{wrapfigure}
The overall approach is depicted on the right: 
our tool \synth generates a passive monitor \mon (written in Scala) from a probabilistic session type $\stS$ that behaves as a partial-identity~\cite{GommerstadtJP:ESOP:18}. 
In addition to carrying out the runtime analysis, such monitors are also tasked with forwarding the messages analysed, offering higher degrees of control for stopping execution once a violation is detected.
Accordingly, the synthesised executable analysis \mon, is instrumented to act as an intermediary proxy between two interacting components participating in the session, \eg between a client (\client) and a server (\srv) for our PST $\stS[game]$ from \cref{ex:monitoring-estimates}. 
Internally, the synthesised monitor \mon uses the \lchannels library \cite{DBLP:conf/ecoop/ScalasY16} to represent the session type within Scala. 
To interact with the components, the monitor makes use of user-supplied \emph{connection managers} (\textsf{CM}) that sit between the monitor and the components.\footnote[1]{%
This design is conveniently inherited from \cite{DBLP:conf/forte/BurloFS20,DBLP:conf/ecoop/BurloFS21} 
but is orthogonal to our approach.
}
The connection managers act as \emph{translators} and \emph{gatekeepers} by transforming messages from the transport protocols supported by \client and \srv to the session type representation used by \mon, and \emph{vice versa}. 
These allow the monitor synthesis to abstract over the communication protocols in use: \ie the synthesis is agnostic to the message transportation being used.

The quantitative analysis of the communicated messages applies only if the qualitative aspects of the type are being respected. 
Similarly to the monitors produced in \cite{DBLP:conf/forte/BurloFS20}, the code synthesised by our tool can be seen as communicating finite-state machines \cite{DBLP:journals/jacm/BrandZ83} where states correspond to choice points in a session type. 
Upon receiving a message, the actual direction of the choice point triggers the analysis of a state's transition modelling such a choice, potentially producing a warning (or a warning retraction) as a side-effect. 
\Cref{alg:synthesised-moitor-state} outlines the logic inside a single state representing a choice point $\stS[j]$. 
The synthesis in \cite{DBLP:conf/forte/BurloFS20} generates the code that conducts the dynamic typechecking on the messages received, and issues violation verdicts committed to $\stS$ (line 12).
In this work, we augment the synthesis to equip the monitor logic with the ability to conduct the quantitative analysis (lines 3 to 8) as discussed in \Cref{sec:methodology}.

\begin{algorithm}[H]
  \label{alg:synthesised-moitor-state}
  \SetKwFor{ForEach}{forall}{calculate}{endfall}
  \SetKwProg{rcv}{receive}{}{}
  \rcv{\normalfont{choice $i$ at choice point $\stS[j]$}}{
    \If{$i \in$ choices $I_j$}{
      increment counters $c_j$ and $c_{i,j}$\\
      \ForEach{choices $i \in I_j$}{
        \If{\normalfont{\textbf{not}} \textit{checkInterval($c_i$, $c_{i,j}$, $p_{i,j}$)}}{
          issue a warning blaming the sender
        }
        \Else{
          retract warning
        }
      }
      \textbf{forward} choice $i$ to the other side\\
      proceed according to the continuation of choice $i$ in choice point $j$ of $\stS$
    }
    \Else{
      issue a violation verdict
    }
  }
  \caption{Synthesised state of a monitor}
\end{algorithm}

The monitors generated by our tool include message counters along with the logic necessary to estimate the probabilities of the choices in a running session. 
This logic is used to issue warnings whenever the observed (partial) execution deviates from the probabilities in each state $\stS[j]$. 
If the message received by the monitor respects the choice point $\stS[j]$ (line 2 in \Cref{alg:synthesised-moitor-state}), it increments the counters of the current choice point, $c_j$, and that of the choice taken $c_{i,j}$. 
For every choice within the choice point, the monitor invokes the function \textit{checkInterval} (described in \Cref{alg:calc-interval}) to \emph{calculate} and \emph{test} whether the estimated probability falls within the respective confidence interval (line 5). 
If the estimated probability is not within the interval, the monitor issues a warning and assigns blame to the sender of the current message (line 6).
Otherwise, if the probability lies within the interval, the monitor retracts a previously issued warning (line 8). 
In an effort to minimise unnecessary (repeated) notifications, the monitors generated by our tool only issue (resp. retract) a warning the \emph{first time} an estimated probability transitions outside (resp. inside) the calculated confidence interval. All subsequent notifications are suppressed in case of estimated probabilities that \emph{remain} outside (resp. inside) the interval.

\Cref{alg:calc-interval} is the function implementing the methodology outlined in \Cref{sec:methodology}. 
It calculates the estimated probability and confidence interval around the probability specified in the type, based on the counters maintained by the monitor itself. 
We note that this function can be adapted to other techniques that test whether the behaviour is being respected without affecting the main synthesis of the monitor; see \Cref{sec:discussion} for further discussions on this point.

\begin{algorithm}[H]
  \label{alg:calc-interval}
  \SetKwProg{Fn}{def}{:}{}
  \Fn{checkInterval($c_i$, $c_{i,j}$, $p_{i,j}$)}{
    \textbf{calculate} the estimated probability $\widehat{p_{i,j}}$ \eqref{eqn:prob-fun} using $c_i$ and $c_{i,j}$\\ %
    \textbf{calculate} the error $E_{i,j}$ \eqref{eqn:err-fun} using $c_j$ and $p_{i,j}$\\
    \KwRet $\bigl(\widehat{p_{i,j}}$ \textbf{in} $[p_{i,j}-E_{i,j}, p_{i,j}+E_{i,j}]\bigr)$
  }
  \caption{Function to calculate intervals}
\end{algorithm}

\begin{example}
  \label{eg:mon-sim1}
  Recall the PST $\stS[game]$ from \Cref{ex:intro} and assume that the monitor synthesised from $\stS[game]$ is instantiated with the confidence level $\confidence = 99.999\%$. 
  Consider the extended case from \Cref{ex:monitoring-estimates} whereby, after guessing incorrectly 4 times, the client asks for \lab{Help} 13 times; 
  it then guesses correctly 2 consecutive times, which brings the monitor counters of the respective internal choice point ($\stSelOp$) to: 

  \smallskip\centerline{\(
    c_{\stSelOp} = 6 \qquad\qquad c_{\lab{Correct},\stSelOp} = 2 \qquad\qquad c_{\lab{Incorrect},\stSelOp} = 4
  \)}\smallskip

  Following the logic in \Cref{alg:synthesised-moitor-state}, after having incremented the counters (line 3), the monitor checks the confidence intervals for \emph{every} choice present in the choice point (line 4). 
  It invokes the function \textit{checkInterval} with the arguments $c_{\stSelOp}$, $c_{\lab{Correct},\stSelOp}$, $p_{\lab{Correct},\stSelOp}$ where $p_{\lab{Correct},\stSelOp} = 0.01$ for the choice \lab{Correct} in $\stS[game]$. 
  The monitor calculates:

  \smallskip\centerline{\(
    \widehat{p_{\lab{Correct},\stSelOp}} = 0.33 \qquad\; E_{\lab{Correct},\stSelOp} = 0.18 \qquad\; p_{\lab{Correct},\stSelOp} \pm E_{\lab{Correct},\stSelOp} = [-0.17,0.19]
  \)}\smallskip
  
  \noindent Since $\widehat{p_{\lab{Correct},\stSelOp}}$ is not included in the interval, the function returns \texttt{False}. 
  Consequently, the monitor issues a warning blaming the server for sending \lab{Correct} with a probability higher than that specified in $\stS[game]$. 
  Next, the monitor invokes \textit{checkInterval} for the choice \lab{Incorrect} with the arguments $c_{\stSelOp}$, $c_{\lab{Incorrect},\stSelOp}$, $p_{\lab{Incorrect},\stSelOp}$ where $p_{\lab{Incorrect},\stSelOp} = 0.99$. 
  Similarly, the monitor calculates:

  \smallskip\centerline{\(
    \widehat{p_{\lab{Incorrect},\stSelOp}} = 0.67 \quad\; E_{\lab{Incorrect},\stSelOp} = 0.18 \quad\; p_{\lab{Incorrect},\stSelOp} \pm E_{\lab{Incorrect},\stSelOp} = [0.81,1.17]
  \)}\smallskip
  
  \noindent and since $\widehat{p_{\lab{Incorrect},\stSelOp}}$ does not lie within the interval, the monitor issues another warning, again blaming the server for this choice. 

  Consider now the case where the client sends 6 further guesses, to which the server replies with \lab{Incorrect} for all. 
  Therefore, the monitor counters for this choice point are now updated as follows: 

  \smallskip\centerline{\(
    c_{\stSelOp} = 12 \qquad\qquad c_{\lab{Correct},\stSelOp} = 2 \qquad\qquad c_{\lab{Incorrect},\stSelOp} = 10
  \)}\smallskip

  \noindent Similarly, the monitor calculates the intervals for both choices: 

  \smallskip\centerline{\(
    \widehat{p_{\lab{Correct},\stSelOp}} = 0.17 \qquad\; E_{\lab{Correct},\stSelOp} = 0.13 \qquad\; p_{\lab{Correct},\stSelOp} \pm E_{\lab{Correct},\stSelOp} = [-0.12,0.14]
  \)}\smallskip
  \centerline{\(
    \widehat{p_{\lab{Incorrect},\stSelOp}} = 0.67 \quad\; E_{\lab{Incorrect},\stSelOp} = 0.13 \quad\; p_{\lab{Incorrect},\stSelOp} \pm E_{\lab{Incorrect},\stSelOp} = [0.86,1.12]
  \)}\smallskip

  Note that after the monitor observes more messages for this choice point, the intervals shrink for both choices, gradually converging to the specified probabilities. 
  Nonetheless, the estimated probabilities from the observed behaviour still do not fall within the confidence intervals in both cases. 
  Accordingly, the monitor does not issue any warnings since it had already issued one previously when the potential violation was originally detected. 
  
  In fact, the monitor only \emph{retracts} the warnings when the estimated probabilities fall within the interval. 
  Concretely, the client would have to guess incorrectly 6 more times, setting the counters to: 

  \smallskip\centerline{\(
    c_{\stSelOp} = 18 \qquad\qquad c_{\lab{Correct},\stSelOp} = 2 \qquad\qquad c_{\lab{Incorrect},\stSelOp} = 16
  \)}\smallskip
  
  \noindent These counters result in the intervals: 

  \smallskip\centerline{\(
    \widehat{p_{\lab{Correct},\stSelOp}} = 0.11 \qquad\; E_{\lab{Correct},\stSelOp} = 0.1 \qquad\; p_{\lab{Correct},\stSelOp} \pm E_{\lab{Correct},\stSelOp} = [-0.09,0.11]
  \)}\smallskip
  \centerline{\(
    \widehat{p_{\lab{Incorrect},\stSelOp}} = 0.89 \quad\; E_{\lab{Incorrect},\stSelOp} = 0.1 \quad\; p_{\lab{Incorrect},\stSelOp} \pm E_{\lab{Incorrect},\stSelOp} = [0.89,1.09]
  \)}\smallskip

  \noindent that both include the estimated probability, causing the monitor to retract the warnings. 
  \exqed
\end{example}

It is often the case that warnings for certain branches have little significance. 
For instance, in \Cref{eg:mon-sim1} above, the monitor also issues a warning for the choice \lab{Incorrect} in addition to that for \lab{Correct}. 
In practice, one might only be interested in knowing that the server replied \lab{Correct} with a \emph{higher} probability than that specified. 
To enable such specifications, we enrich the syntax of the probabilistic session types in \Cref{sec:methodology} with the possibility of using $\ast$ which specifies to the monitor to not issue warnings for the respective branch or interval boundary. 

\begin{example}
  The PST $\stS[game]$ from \Cref{ex:intro} can be modified to the type description  $\textsf{S}'_\textsf{game}$ below: 
  \begin{align*}
    \textsf{S}'_\textsf{game} = \stRec{X}.\stBraOp
    \left\{\begin{array}{l}
          \stRcv{Guess}{\stVarDec{num}{\typeInt}}{0.75,*}.
          \stSelOp\left\{\begin{array}{l}
                    \stSnd{Correct}{}{0.01}.\stRecVar{X},
                    \\
                    \stSnd{Incorrect}{}{*}.\stRecVar{X}
                    \end{array}\right\},
          \vspace{.2em}\\
          \stRcv{Help}{}{*,0.2}.\stSnd{Hint}{\stVarDec{info}{\typeStr}}{*}.\stRecVar{X},
          \vspace{.2em}\\
          \stRcv{Quit}{}{*}.\stEnd \end{array}\right\}
  \end{align*}
  The new type indicates to the monitor \emph{exactly} which choices it should issue warnings for. 
  For the external choice, the monitor should only issue a warning when the estimated probability of the client sending \lab{Guess} is \emph{lower} than 0.75 and that of sending \lab{Help} is \emph{higher} than 0.2, completely ignoring the probability of the choice \lab{Quit}. 
  Similarly, for the internal choice, the monitor should only issue a warning for the choice \lab{Correct}, and suppress those for \lab{Incorrect}. \exqed
\end{example}

With this minor extension to the probabilistic session types we reduce the number of warnings issued and retracted by the monitor. 
Moreover, we also decrease the amount of computation performed by the monitor at runtime to only those choices that are deemed important. 
Effectively, this improves the overheads induced by the monitor. 
\section{Conclusions and Discussion}\label{sec:discussion}

We have presented a tool-based methodology to analyse specifications augmented with quantitative requirements \emph{at runtime}. 
More specifically, we extend existing work to implement the synthesis of monitors from probabilistic session types that conduct analysis of the interaction between two parties at runtime. 
The synthesised monitors issue warnings based only on evidence observed up to the current point of execution while taking into account its accuracy. 
Notably, the proposed methodology can serve as the basis for other runtime analysis techniques in which the specifications describe any quantitative behaviour. 

We conjecture that our approach can be used for systems where protocol-based interactions are replicated in large numbers, and where human intervention is required to ensure their correct execution (\eg healthcare and fraud detection in e-payments or online gambling).
In such applications, our monitors would direct human operation (\ie the scarce resource) to
the cases that have the highest likelihood of exhibiting anomalous behaviour. 
Another potential application is that of control software that is derived using AI learning techniques.
Although effective, such software is often not fully understood and notorious for sudden unexpected behaviour.
With our approach, we can automate the monitoring of its interactions and shut off communication whenever the approximated runtime behaviour deviates considerably from that projected.

\subsection{Related Work}
Our methodology uses PSTs akin to those introduced in \cite{imptt20}. 
The authors use a type system to \emph{statically} estimate the probability of well-typed processes 
\emph{reaching successful states}. 
Notably, types are dynamically checked in our approach and we do not guarantee probabilistic properties; the proposed runtime analysis only issues warnings when the observed behaviour at runtime substantially deviates from the specification. 
Moreover, our PSTs can also specify behaviour of deterministic systems and are not restricted to probabilistic systems. 
Several works apply probabilistic monitoring to minimise the number of runs to be monitored, based on predefined confidence levels \cite{grunske2011effective,zhu2013bayesian,ruchkin2020compositional}. 
In \cite{stoller2011runtime}, probability is estimated to check whether the system's behaviour modelled as a hidden Markov model satisfies a temporal property in cases where gaps are present in the execution trace. 
Unlike these approaches, we use probabilities to specify quantitative aspects of communication protocols which we then check whether they are being respected at runtime.

On the runtime verification of probabilistic systems, the work in \cite{DBLP:journals/tse/FilieriTG16} models systems as discrete-time Markov chains and expresses requirements using Probabilistic Computation Tree Logic. 
Their aim is to adapt the behaviour of the underlying system to satisfy non-functional requirements, such as reliability or energy consumption. 
The work in \cite{Esparza2020:techrep} also monitors Markov Chains whereby monitors are able to verify if a property is satisfied by executing the system and steer it to take certain paths. 
Similarly to \cite{DBLP:journals/tse/FilieriTG16}, Markov decision processes are used in \cite{forejt2012incremental} to model probabilistic systems and optimise the performance of their verification with the aim of using the results obtained to steer the system. 
In \cite{DBLP:journals/tse/FilieriTG16}, \cite{Esparza2020:techrep} and \cite{forejt2012incremental}, the authors adopt incremental verification techniques that exploit the results of previous analyses of the system, whereas our runtime analysis only considers the \emph{current} execution without any prior knowledge on the system.
Moreover, we employ monitors that are passive and do not alter nor control the behaviour of the monitored system in any way. 
\subsection{Future Work}

\paragraph{Improving confidence interval estimation.}
The proposed approach using on confidence intervals described in
\Cref{sec:methodology} serves the goal of instantiating our
interpretation of probabilistic session types (\Cref{sec:prob-session-types})
on a concrete mechanism for monitors to emit judgements on
the probabilistic behaviour of components at runtime.
Our approach is not limited, nor bound, to the current statistical inference technique. 
For instance, we can improve our CI estimation by utilising the Wilson
score interval \cite{Wilson1927}, which is more costly but also more
reliable than normal approximation when the sample size (observed
messages up to the current point of execution) is small or the
specified probability is close to 0 or 1.
With an easy extension, we can additionally support different confidence levels per choice points.
We also plan to study alternative statistical inference paradigms apart from the frequentist approach considered here, such as the Bayesian intervals \cite{DBLP:journals/sma/AgrestiH05} which would potentially give different interpretations to the PSTs. 
In turn, this would enable monitors to issue warnings based on the \emph{aggregate recommendations} of the different estimators. 

\paragraph{Alternative interpretations of probabilistic session types.}

In the proposed approach we opted for an interpretation of probabilistic session types that only considers probabilities at each individual choice point rather than the \emph{global probabilistic behaviour}.
We plan to study different interpretations that consider the \emph{dependencies} among choice points in a session type, and also probabilities that are data (payload) dependent. 

An intricate aspect of our approach is that it
may lead to a potential trade-off between two extremes: taking longer to issue a warning with high confidence, or issue warnings earlier with low confidence. 
The first extreme corresponds to an increased risk of false negatives: a monitored application could substantially diverge from the observed session type, without being flagged.
The second extreme corresponds to an increased risk of false positives: a statistically well-behaving monitored application could be flagged after a minor divergence from the expected frequency of choices.
Finding the right value for the confidence level requires careful calibration since it depends on the application and on the relative cost of false negatives when compared to that of false positives.

Application-dependent heuristics together with more advanced interpretations of probabilistic session types can be used to overcome these difficulties. 
For instance, we could consider the introduction of an observation window of length $w$, and the probabilities in the session type could refer to a limited number of observed communications regulated by $w$. %
Accordingly, the monitor could keep track of the number of communications observed in the desired window, and use this information to issue warnings in case deviations persist.
Concretely, with the interpretation of probabilities described in \Cref{sec:prob-session-types}, 
a client in \Cref{ex:intro} is allowed to send any number of consecutive requests for help, as long as their frequency is not far from 20\%, calculated by considering the \emph{entire history} of the session.
By introducing a small observation window, we can ensure that two consecutive requests of help would be flagged at any point during the session, regardless of their frequency.

\paragraph{Other extensions and improvements.}
We are in the process of conducting empirical evaluations to assess the effectiveness of our methodology. 
There are a number of extensions that can be realised relatively easily to improve the tool's flexibility and applicability.
For instance, our methodology can be extended so that the probabilities within the session type are (machine-)learnt from a series of observed interactions of different parties.
Moreover, monitor (verdict) \emph{explainability}~\cite{Francalanza21,DawesR19} is rapidly gaining importance: our tool can be readily extended to provide useful explanations to support the warnings raised.
We are also considering the adaptation of our methodology to the pre-deployment phase of development, thus turning monitors into test drivers that can steer-and-verify implementations.

\bibliographystyle{splncs04}

\bibliography{main}

\end{document}